\documentstyle[twoside]{europhys}

\def\stars{\bigskip\centerline{***}\medskip}

\newif\ifboo \boofalse

\def\Review#1{\boofalse{\it #1},}
\def\Name#1{{\sc #1},}
\def\Vol#1{\ifboo Vol. {\bf #1}\else{\bf #1}\fi}
\def\Year#1{\ifboo #1\else(#1)\fi}
\def\Book#1{\bootrue{\it #1},}
\def\Page#1{\ifboo {\rm p. #1}\else{\rm #1}\fi}

\input epsf

\begin{document}

\euro{}{}{}{} \Date{} 

\title{Temperature cycling during the aging of a polymer glass}

\author{L. Bellon, S. Ciliberto, C. Laroche}
\institute{
    Laboratoire de Physique - U.M.R. 5672 \\
     \'Ecole Normale Sup\'erieure de Lyon \\
     46, all\'e d'Italie, 69364 Lyon Cedex 07, France}


\pacs{ \Pacs{77}{22Gm}{Dielectric loss and relaxation}
\Pacs{64}{70Pf}{Glass transitions}\Pacs{05}{20$-$y}{Statistical
mechanics}}

\maketitle

\begin{abstract}
Temperature cycling is used to study aging properties of
plexiglass (PMMA) dielectric constant. If a negative temperature
cycle is applied during the aging time the relaxation dynamics is
just delayed for a time equal to the cycle period. In contrast
this time shifting procedure cannot be applied to positive
temperature cycles. The analogies and differences with similar
experiments done in other aging systems, such as spin glasses,
orientational glasses and supercooled  liquids, are discussed.
\end{abstract}

\section{Introduction}
It is very well known  that amorphous materials such as polymer
solids present long relaxation time, when they are rapidly cooled
from the liquid phase to any temperature below their glass
transition temperature $T_g$ \cite{Struick}. This phenomenon which
is known as aging has been widely studied  in spin-glasses (SG),
where remarkable theoretical and experimental progress have been
achieved \cite{Bouchaud, Vincent}. In these systems many features
of the relaxation dynamics have been understood by imposing
temperature cycles to the samples. The response of the systems to
these cycles has been interpreted with a hierarchical structure of
the free energy landscape, which evolves with temperature
\cite{Vincent, Hammann}, or in a mean-field theoretical approach
\cite{Kurchan}. Very recently such a kind of experiments have been
repeated in supercooled liquids (SL) \cite{Nagel} and in
orientational glasses (OG) \cite{Alberici} by measuring the
dielectric susceptibility of the sample.
 The relaxation dynamics
turned out to be rather different from that of spin-glasses, and
 a domain growth
interpretation \cite{Doussineau} seems to be rather appropriate
for OG and SL. Temperature cycling has been also  applied in the
study of polymer mechanical properties \cite{Struick,Cavaille} but
as far as we know no direct comparison to the previous systems has
been done.

The purpose of this letter is just to describe a series of
experiments where temperature cycling has been applied to the
measurement of the dielectric constant of polymers, specifically
of PMMA. The dielectric constant measurement is a common method to
analyze aging properties of polymers. After presenting the results
of our measurements, we will discuss differences and similarities
between polymers and other aging systems.

\section{Experiment}
To determine the dielectric constant, we measure the complex
impedance of a capacitor whose dielectric is the PMMA sample. In
our experiment a disk of PMMA of diameter $10cm$ and thickness
$0.3mm$ is inserted between the plates of a capacitor whose vacuum
capacitance is $C_o=230 pF$. The capacitor  temperature has a
stability is $0.1 K$ and it may be changed from $300 K$ to $500K$.
Temperature quenches  can be done at a rate of $3 K/min$.

The capacitor is a  component   of the feedback loop of a
 precision  voltage  amplifier whose input is connected
to a signal generator. We obtain the real and imaginary part of
the capacitor impedance by measuring the response of the amplifier
to  either to a white-noise or to a sinusoid.  This apparatus
allows us to measure the real and imaginary part of dielectric
constant $\epsilon=\epsilon_1 +i \ \epsilon_2$
 as a function of temperature $T$, of frequency $\nu$
and time $t$. The sensitivity of the measurement system is such
that $\epsilon_2$ as small as $4 \ 10^{-3} \epsilon_1$ can be
measured at a frequency of $0.1Hz$. We use this apparatus to study
the aging properties of $\epsilon$  for  $0.1 Hz< \nu <20Hz$ and
$310K<T<410K $. The glass transition temperature for PMMA is
$T_g\simeq 388K$.  The $\alpha$ and $\beta$ relaxation frequencies
at $T_g$ are for PMMA $f_\alpha=10^{-3}Hz$ and $f_\beta\simeq 2 \
10^4 Hz$\cite{book}. These two frequencies, which are a decreasing
function of $T$,  remain well outside the frequency range used in
this experiment for $310K<T<T_g$. Therefore we probe the aging of
$\epsilon$ well above $f_\alpha$ and well below $f_\beta$. The
following discussion will focus on the imaginary part of the
dielectric constant $\epsilon_2$, but we could have chosen
$\epsilon_1$ as well as its behavior leads to the same results.

\begin{figure}
\begin{center}
\epsfbox{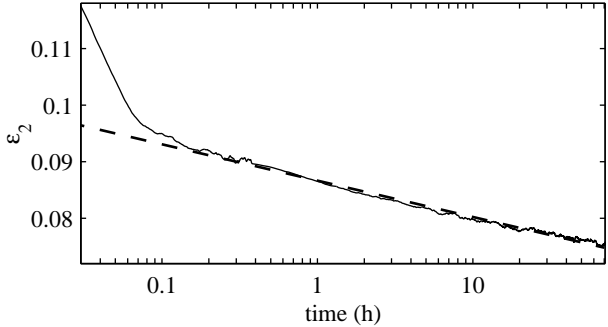}
\end{center}
\caption{ Dependence on $t$ of $\epsilon_2$, measured at $\nu=1Hz$
after a quench at $T_1=365K$. The logarithmic fit is accurate over
 two decades.
 }
\label{fig:refagi}
\end{figure}

The measurement is performed in the following way. The sample is
before heated  till  a temperature  $T_o=410K$ which is higher
than $T_g$. After having left the sample at $T_o$ for a few hours
the temperature is rapidly decreased  in about $15 min$ at a
temperature $T_1<T_g$. The temperature is then regulated by the
oven. The zero of the aging time is taken at the instant, during
the quench, when the sample temperature is equal to $T_g$. A
typical aging curve at $1Hz$ and $T_1=365K$ is shown in
fig.\ref{fig:refagi} where $\epsilon_2$ is plotted as a function
of time. We clearly notice a logarithmic dependence on time of the
dielectric constant after the first $10 min$ from the quench. The
logarithmic dependence on time of $\epsilon$ is observed in all
the frequency range we have explored.

Repeating this experiment for various
 $T_1$ and $\nu$ we find that the
 aging rate depends  on this temperature and on frequency. The
sample properties evolve faster when $T_1$ is close to $T_g$.
Specifically one can write
$\epsilon_j(T_1,t,\nu)=A_j(T_1,\nu)-B_j(T_1,\nu) \ \log(t/t_o)$
with $j=1,2$ and $t_o=1h$. Here $A_j$ is the value of $\epsilon_j$
at $t=t_o$
 It is
found that $A_j$ and  $B_j$ are functions of $T_1$ and of the
frequency $\nu$ at which the dielectric constant is measured
\cite{Remarque1}.

\begin{figure}

\centerline{ \epsfbox{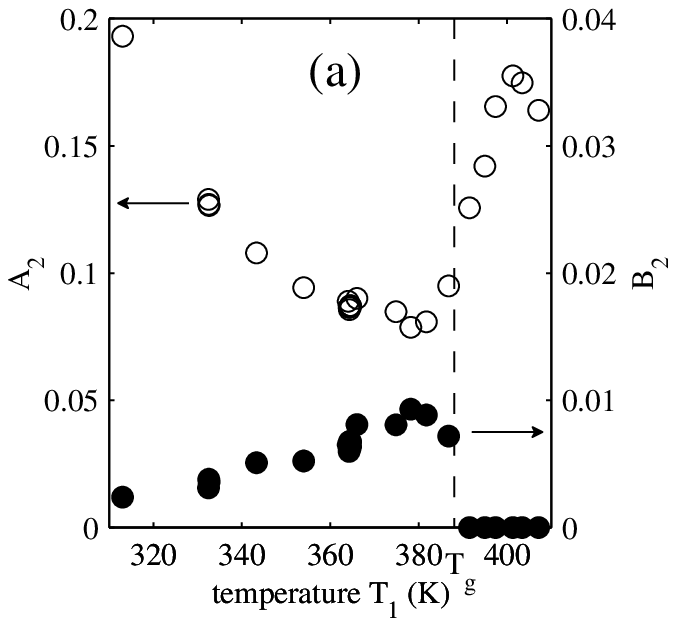} \hspace{5mm}
\epsfbox{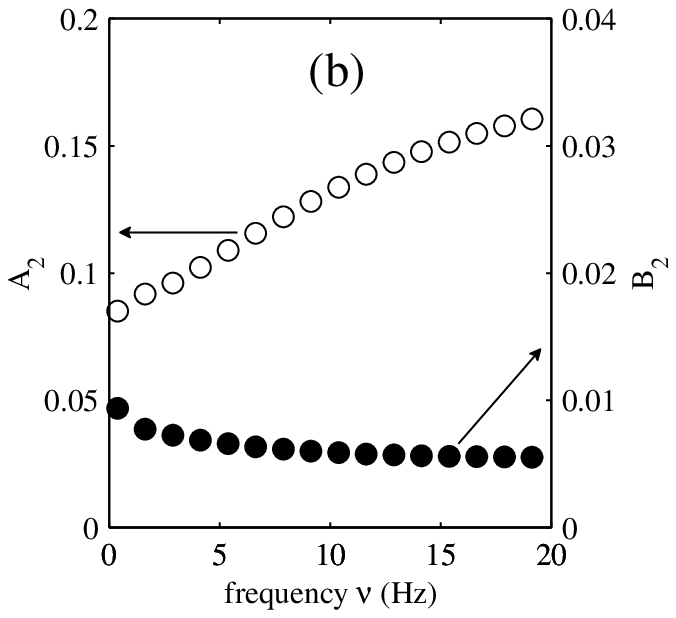}}

 \caption{
(a) Dependence on $T_1$ of $A_2 \ (\circ)$ and $B_2 \ (\bullet)$
at $\nu=1Hz$. (b)  Dependence on $\nu$ of $A_2 \ (\circ)$ and $B_2
\ (\bullet)$ at $T_1=366K$.
 }
\label{fig:alpha}
\end{figure}

The values of $A_2$ and  $B_2$, measured at $\nu=1Hz$, are plotted
in fig.\ref{fig:alpha}(a) as a function of $T_1$.
The imaginary part of
$\epsilon$ increases at low and high temperature. The high
temperature peak at about $400K$ is related to the $\alpha$
relaxation ($f_\alpha\simeq 1Hz $ at $400K$), whereas the increase
at small temperature is related to the $\beta$
relaxation\cite{book}.
 Notice that $B_2$ is an
increasing function of $T_1$ till a temperature $T_m$ close to
$T_g$, and then goes down to 0 if $T_1>T_g$. Indeed for a quench
temperature larger than $T_g$ the sample can reach a thermodynamic
equilibrium, so no aging is observed but only an exponential decay
of $\epsilon$ toward its stationary value.

The values of $A_2$ and  $B_2$, measured at $T_1=366K$, are
plotted in fig.\ref{fig:alpha}(b) as a function of $\nu$. The
dependence on $\nu$ of $A_2$ is very weak in this frequency range.
No relaxation peak is observed indeed $f_\alpha \simeq 10^{-7} Hz$
and $f_\beta \simeq 15 kHz $ at $366K$. $B_2$ is a  slowly
increasing funtion for $\nu\rightarrow 0$. Indeed aging is smaller
at high frequencies than at low frequencies (a theoretical
justification of such a behaviour can be found for example in
ref.\cite{Bouchaud}).

In order to check for memory effects during the aging process, we
have  submitted  the sample to temperature jumps around $T_1$. The
following procedure has been used. After the quench from $T_o$ to
$T_1$ the temperature is maintained at $T_1$ for $t_a=2h$, and
then suddenly changed to $T_2$. The sample temperature is
regulated at $T_2$ for a time $\delta t=4h$ and at time
$t_b=t_a+\delta t$ once again changed to $T_1$. We will present
separately the two cases $\Delta T=T_2-T_1<0$ and $\Delta T > 0$.
 We  show the results of the experiment at $\nu=1Hz$
 and  $T_1=365K$  because this set of values give the best experimental
  accuracy in the analysis of the response of the system to a
  temperature perturbation. Indeed, looking at
  fig.\ref{fig:alpha},
  one immediately sees that at this
temperature and this frequency $B_2$ is large enough to have a
large variation  in a short time. Further $T_1$ is sufficiently
far away from $T_g$ to make positive and negative cycles with the
same amplitude, remaining always below $T_g$. The experiment have
been done at other values of $T_1$ and $\nu$ but no  differences
on the response of the system to the temperature perturbation  are
observed.

\begin{figure}
\centerline{ \epsfbox{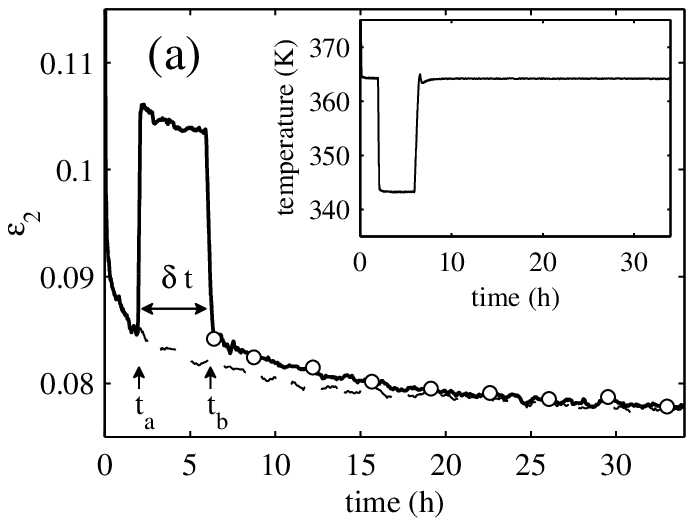} \epsfbox{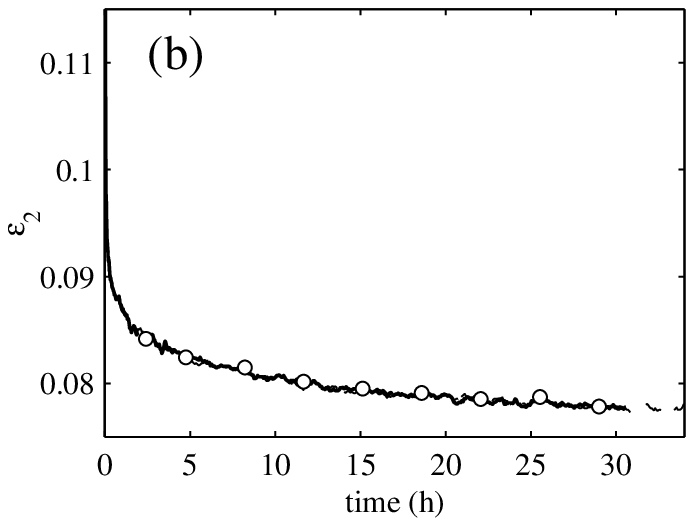}}

\caption{ (a) A temperature perturbation $\Delta T = -22K$ is
applied at $t_a=2h$ and switched off at $t_b=6h$ (see inset : time
history of the sample temperature). The corresponding time
evolution of $\epsilon_2$ at $\nu=1Hz$ (continuous line) is
compared with the standard relaxation curve (dashed line) at
$T_1=365K$. The symbols ($\circ$) mark the evolution after the
temperature perturbation (i.e. for $t>t_b$).  (b) Same curves of
a), but the points of the aging curve after the perturbation
($-\circ-$ line) are now shifted  of $\tau=-4h$ in order to
superpose them with the standard aging curve. }

\label{fig:cycle1}
\end{figure}

Negative cycle ($\Delta T<0$): a typical time history of the
sample temperature is plotted in the inset of
fig.\ref{fig:cycle1}(a), with $T_1=365K$ and $T_2=343K$. The
corresponding aging of $\epsilon_2$ is compared in the figure with
the standard aging curve at $T_1$. We clearly see that for
$t>t_b$, when the temperature comes back to $T_1$ the
corresponding aging curve is different from the standard one at
$T_1$. However  in fig.\ref{fig:cycle1}(b) we show that it is
possible to superpose the two curves, by applying a time shift
$\tau=-4h$ to the values of $\epsilon_2$ measured after the cycle
at $T_2$ (i.e. for   $t>t_b$). We see that after the shifting
procedure \cite{fit} the time behavior is the same with and
without temperature modulation: there is a memory effect of the
previous history of the sample.

\begin{figure}
\centerline{ \epsfbox{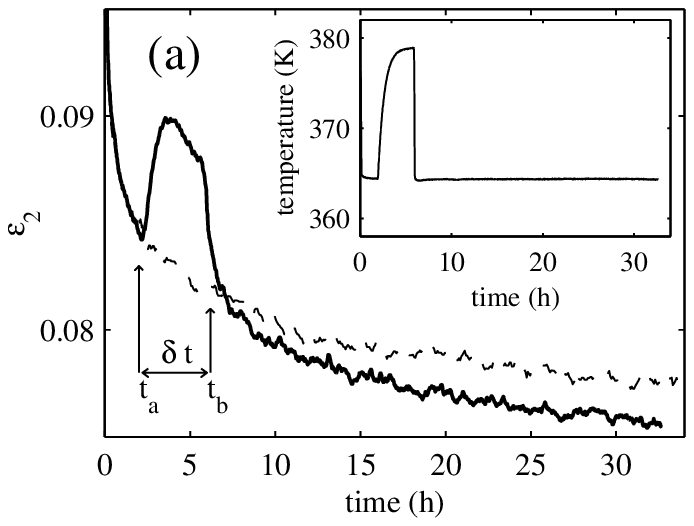} \epsfbox{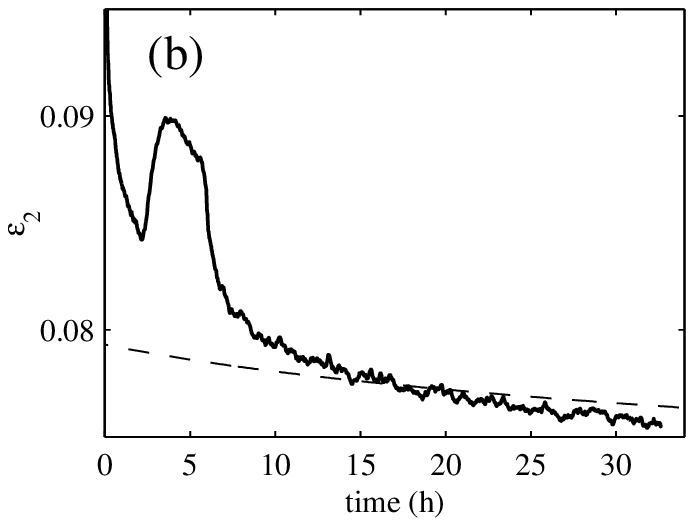}}

\caption{ (a)  A temperature perturbation $\Delta T = +15K$ is
applied at $t_a=2h$ and switched off at $t_b=6h$ (see inset : time
history of the sample temperature). The corresponding time
evolution of $\epsilon_2$ at $\nu=1Hz$ (continuous  line) is
compared with the standard reference relaxation curve (dashed
line) at $T_1=365K$ (b) The perturbed aging curve (continuous
line) is compared with the shifted reference curve (dashed  line)
with $\tau=16h$ (see \cite{fit}).
 The time shifting
procedure does not work this time.} \label{fig:cycle2}
\end{figure}

As suggested by the behavior of $B_2$ for low temperatures, aging
is slower at $T_2$ than at $T_1$ : we notice that for this
experiment the time shift $\tau$ is very close to $-\delta t$, as
if nothing had happened during the cooler period. In fact we
expected $\tau$ to be a function of $\Delta T$, with $\tau
\rightarrow 0$ when $\Delta T \rightarrow 0$. So we repeated this
temperature cycling experiment for various $\Delta T$. The
surprising result of these measurements is that for all
perturbations ($T_1=365K$ and $\Delta T=-32, -30, -22, -12$ and
$-5K$) except the smallest one, $\tau \simeq -\delta t \cdot (1\pm
0.1) $ and the only evidence of $\tau$ vanishing to zero is $\tau
= - 0.6 \delta t$ for $\Delta T = -5K$. The systems acts as if
aging at any temperature $T_2$ a bit below $T_1$ during the cycle
was almost useless when back to $T_1$.

Positive cycle ($\Delta T >0$) : the results of a measurement with
$\Delta T>0$ are shown in fig.\ref{fig:cycle2}. If we were to
define a shifting time $\tau$ for this experiment, we would expect
it to be positive as aging is faster at higher temperature.
Indeed, the values of $\epsilon_2$ after the cycle are under the
standard aging curve in fig.\ref{fig:cycle2}(a). Thus they
correspond to an older reference sample. But as one can see in
fig.\ref{fig:cycle2}(b), the best time shift of the experimental
data  \cite{fit}  to superpose it to the standard aging curve
doesn't lead to a good approximation.

A positive cycle cannot be summarized by a time shift : if aging
has been accelerated by the temperature perturbation, its behavior
after the cycle is not equivalent to a redefinition of the time
origin for the standard curve.

\section{Comparison with other temperature cycling experiments}

The results of these temperature cycling measurements are rather
different from those of SG \cite{Vincent}, or SL \cite{Nagel}
experiments, but quite similar to the OG \cite{Alberici}
measurements. Let us study separately again the two cases $\Delta
T < 0$ and $\Delta T > 0$.

When a negative temperature cycle is applied to any of the former
systems, the common trend for long time after the cycle is a delay
of the aging process. For both SG and OG, a backward time shift
$-\tau \leq \delta t$ can be applied to the aging curve after the
cycle in order to superpose it with the standard aging curve,
exactly as in our experiment. But for both systems, the short time
behavior after the perturbation had to be discarded since a fast
contribution to the aging dynamics gives an overshoot of the aging
curve. This effect, which can be understood in a mean-field theory
framework for SG \cite{Kurchan}, is absent or at least really weak
in PMMA.

In SL, this ``short time'' behavior (undershoot of the aging curve
this time) is in fact dominant for a time of the same order of the
perturbation duration. That's why Leheny and Nagel \cite{Nagel}
give no estimation of $\tau$ since the delay  of the aging process
for long times is not the main consequence of the cycle. Again
PMMA is very different since the time shift applies to any time $t
> t_b$.

When a positive temperature cycle is applied to SG or OG, these
systems tend to ``forget'' their previous history (no data of
positive cycle for SL is available to our knowledge).  PMMA
definitely doesn't behave as SG for positive cycle, but  it is
quite close to OG, where the long time behavior would corresponds
to a different cooling rate (and therefore a different final
equilibrium state) rather than a different time origin. Indeed we
checked that a slower cooling leads to smaller values of
$\epsilon_2$ so that for the same age a slowly cooled sample looks
older than a rapidly quenched one.  The relative amplitude of this
rate dependence is comparable  to the one of OG
\cite{Remarque1,rate}.

\section{Conclusion}
We have described the results of  a temperature cycling experiment
on the aging of the dielectric constant of PMMA.
 We have shown that  in PMMA, as in OG and SG,
 the  dynamics is quite different for positive
and  negative temperature cycles. The negative pulses produces
just a time delay in the evolution whereas the positive pulses
have the tendency to reinitialize the aging. The main difference
between the OG and SG dynamics and that of the PMMA is in the
 the  presence of an overshoot after the negative
cycles in  OG and SG. These overshoots are absent in PMMA.
 Furthermore in PMMA the
values of $\epsilon$ strongly depend on the cooling rate as in the
OG, which is not the case for SG.

The main question is to find the more suitable model which can
describe these observations. On a first approximation our  results
could be qualitatively explained by the aging models
\cite{Vincent, Hammann} based on a free energy landscape with  a
temperature dependent hierarchical structure. This  rather
qualitative model has been constructed to account for the general
behavior of spin glasses. It seems to be adapted to our
experiment.  The sensitivity of the PMMA free energy landscape  on
temperature can be estimated using the dependence on temperature
of the delay time for the negative temperature cycles. As no
change in $\tau$ has been observed for $\Delta T> 5K$ we can claim
that one should change temperature of at least $5K$ in order to
strongly modify the PMMA landscape and its time evolution.

 Recently two
other models have been proposed \cite{Kurchan,Doussineau}. One
\cite{Kurchan} is based on a mean field approximation of a
continuous spin system \cite{Mezard, Cugliandolo}.  This model
explains quite well the SG dynamics and the presence of overshoot
but as pointed out by the authors it cannot be applied to OG
\cite{Kurchan}.  The second model \cite{Doussineau} is based on
thermally activated domain growth accounts for the experimental
results of OG. Based on the similar cooling rate dependence of the
OG and PMMA dielectric values one could probably conclude that the
model of ref. \cite{Doussineau} is more suitable to describe
 PMMA aging. New insights to this problem should  be given by performing other
 perturbation procedures like those recently applied to OG and SG \cite{memory,bellon}.

\stars
We acknowledge useful discussion with J. Kurchan. This work
has been supported by the "Programme Th\'ematique Mat\'eriaux" de
la R\'egion Rh\^{o}ne-Alpes.

\end{document}

\newpage

PHRASES EFFACEES

(****This behavior is quite similar to the aging of OG  This fast
contribution is not observed in our polymeric sample. **** )

 Another interesting model of
temperature cycling has been recently proposed \cite{Kurchan}. It
is based on a mean field approximation of a continuous spin system
\cite{Mezard, Cugliandolo}. The key point is the decomposition of
the correlation and response functions in a fast and a slow part.
This model explains quite well the SG dynamics and the presence of
overshoot. However it cannot be applied to OG \cite{Kurchan}.
Based on the similar cooling rate dependence of the OG and PMMA
dielectric values one could probably conclude that the model of
ref.\cite{Kurchan} cannot be straightforward applied to PMMA
aging. More precise answer on this point could be given by
measuring experimentally the ratio between correlation and
response functions. Work is in progress in this direction.

For SG, $\Delta T > 0$ leads to a reinitialization of the aging
that was achieved before $t_a$ in the sense that after the cycle,
the system looks younger than it is (which would lead to a
backward time shift $-\tau \sim t_b > \delta t$).

in is constructed by two thick copper plates. Air can circulate
inside the plates in order to cool them very rapidly during the
temperature quench. The temperature of the PMMA sample is measured
by means of two platinum resistance inserted inside the copper
plates. The capacitor  is  placed inside  a metallic screen to
avoid electro-magnetic interferences during measurements. The
capacitor and the screen are inside a temperature stabilized oven,
which has a temperature stability of and a temperature range
extending .

\begin{thebibliography}{}
\bibitem{Remarque1}
In fact, $A_j$ and $B_j$ are also functions of the cooling rate.
In PMMA, $\epsilon_2$ changes of about $25\%$, when the cooling is
changed by two orders of magnitude. More details will be given
elsewhere. All our measurements where performed here with the same
rate of $-3K/min$.
\bibitem{fit} The value of $\tau$ is found by a best fitting procedure
which minimizes the difference for $t>t_b$ between the perturbed
aging curve
 and the shifted reference curve
$\epsilon_2^{ref}(t)=A_2^{ref}-B_2^{ref} \
\log(\frac{t+\tau}{t_o})]$, where $ref$ stands here  for reference
curve.
\end{thebibliography}
\end{document}